\documentstyle[12pt]{article}
\renewcommand{\baselinestretch}{1}
\begin{document}

\begin{titlepage}

\begin{flushright}

BRX TH--431 \\
ULB-TH--98/04\\
WIS/98--56

\end{flushright}

\begin{centering}

{\LARGE $p$-brane Dyons,
$\theta$-terms and Dimensional Reduction}

\vspace{1cm}

{\large S.\ Deser$^{a}$, M.\ Henneaux$^{b,c }$ 
and A.\ Schwimmer$^{d}$}

\vspace{0.5cm}

{\sl
$^a$ Department of Physics, Brandeis University,\\
Waltham, MA 02254, U.S.A.\\[1.5ex]

$^b$Facult\'e des Sciences, Universit\'e Libre de
Bruxelles,\\
Campus Plaine C.P. 231, B--1050 Bruxelles, Belgium\\[1.5ex]

$^c$ Centro de Estudios Cient\'\i ficos de Santiago,\\
Casilla 16443, Santiago 9, Chile\\[1.5ex]

$^d$ Department of the Physics of Complex Systems, 
Weizmann Institute, \\
Rehovot 76100, Israel}

\renewcommand{\baselinestretch}{2}
\small
\normalsize
\begin{abstract}
We present two novel derivations of the recently
established $(-)^p$ factor in the charge quantization
condition for $p$-brane dyon sources in spacetime dimension
$D$=$2p$+2.
The first requires consistency of the condition under
the charge shifts produced by
(generalized) $\theta$-terms.  The second 
traces the sign difference between adjoining dimensions
to compactification effects.
\end{abstract}

\end{centering}

\vspace{0.5cm}

\vfill

\noindent{\tt deser@binah.cc.brandeis.edu; henneaux@ulb.ac.be; 
ftschwim@wicc.weizmann.ac.il}
\end{titlepage}

\pagebreak

\renewcommand{\baselinestretch}{2}
\small
\normalsize
It was recently established \cite{DGHT2}
that the generalized dyon 
quantization condition, for ($p$--1)-brane dyons
coupled to Abelian $p$-forms in 
spacetime dimension $D$=$2p$+2, involves
a $p$-dependent sign:
\begin{equation}
e \bar{g}+(-)^p \bar{e} g = 2 \pi n \hbar, 
\; n \in Z \; .
\label{quantcond}
\end{equation}
The $-$ sign is of course that familiar in $D$=4
electrodynamics \cite{Dirac,Schwinger,Zwan}.  
This sign dependence
was actually anticipated \cite{DGHT1} through analysis of 
chiral sources coupled to chiral $2p$-forms.  It was
particularly stressed in \cite{Bremer}, where
it was related to supergravity duality groups
in higher dimensions \cite{Ferrara}. Another approach
is based on dyon-dyon scattering; 
the relation between $D$=10 and
$D$=4 is also discussed there \cite{Juliaetal}. 

Our aim
is to illuminate this phenomenon through two new arguments.
The first is based on the study of the shift in the ``electric"
charges induced by a ($p$-generalized)
$\theta$-term.  The other uses dimensional reduction, or rather
enhancement, to relate the conditions (\ref{quantcond})
in adjoining dimensions. For concreteness, we shall work
primarily with $D$=4 one-forms and $D$=6 two-forms to
illustrate the generic situations.

\noindent{\bf 1. ~ $\mbox{\boldmath$\theta$}$-terms.}\\
a)~ In $D$=4 it is well known 
that adding a $\theta$-term, $(\theta/2) F_{\mu \nu}\,
^*\!F^{\mu\nu}$ (* always represents dualization)
to the Lagrangian
has the effect of shifting the electric charge of an
$(e,g)$ dyon according to \cite{Witten}
\begin{equation}
e^\prime = e - 2g \theta \; .
\label{shift4D}
\end{equation}
A remarkable feature of this shift is its compatibility 
with the usual Dirac
quantization condition for electric and magnetic charges.  
Namely, if one simultaneously shifts 
all dyon electric charges according
to (\ref{shift4D}) starting from values $(e_a,g_a)$ that
obey (\ref{quantcond}), then the charges
$(e'_a,g_a)$ also do, because
\begin{equation}
e'_a g_b - e'_b g_a = (e_a - g_a \theta) g_b
- (e_b -g_b \theta) g_a 
= e_a g_b - e_b g_a \;. 
\end{equation}
The $-$ sign is crucial in this result.  Indeed, it is
the answer to the converse question: what sign in 
the quantization condition (\ref{quantcond}) leaves it
invariant under the shift
(\ref{shift4D})?\\
b)~ In $D$=6, there
is no $\theta$-term for a single $2$-form since 
$F_{ABC}\, ^*\!F^{ABC}$ vanishes identically.  
However, a $\theta$-term {\it is} possible
with two $2$-forms $A^{(i)}, \; i=1,2$.
The sources here are strings characterized by four 
strengths
(``charges") $(e^{(i)}_a, \; g^{(i)}_a)$, the
respective electric (magnetic) charges of string $a$
coupled to $A^{(i)}$.
We use a uniform convention for the signs of
the couplings $(e_a^{(i)}, \;g_a^{(i)})$ to the 
$2$-forms: the electric couplings enter
with the same sign in the minimal coupling term
$\displaystyle{\sum^2_{i=1,a}} e_a^{(i)} \int A^{(i)}$,
for example.
Single-valuedness of the wave function leads to 
the quantization condition\footnote{One canonical way to
derive this condition consists of attaching
Dirac membranes  -- the higher
dimensional analogs of the Dirac strings \cite{Dirac} --
to the sources \cite{Nepo,Teitel}. Requiring the membranes to
remain unobservable quantum-mechanically then implies
(\ref{quanti6D}).  Indeed, the phase picked
up by the wave-function when the Dirac membrane attached
to string $a$ performs a complete turn around string $b$, 
while the Dirac membrane attached to string $b$ simultaneously
performs a complete turn around string $a$ 
(the ``double-pass" of \cite{DGHT2}) is $(1/\hbar)$ 
times the left-hand side of
(\ref{quanti6D}).}  
\begin{equation}
(e^{(1)}_a g^{(1)}_b \pm g^{(1)}_a e^{(1)}_b)
+ (e^{(2)}_a g^{(2)}_b \pm g^{(2)}_a e^{(2)}_b)
= 2 \pi n \hbar, \; n \in Z
\label{quanti6D}
\end{equation}
with a relative $+$ sign between the contributions 
associated with the two
$2$-forms because of our identical coupling conventions.  
We have left the $\pm$ sign open in (\ref{quanti6D})
to show next how the $\theta$-angle argument selects the 
$+$ sign.  [Of course, for sources that couple to only one
of the fields -- say $A^{(1)}$ --, the second term on
the left is absent in (\ref{quanti6D}).]
Now add the extended $\theta$-term
\begin{equation}
\frac{1}{2} \,\theta \:
\epsilon_{ij} F^{(i)}_{ABC} \,
^*\!F^{(j)ABC} \label{eq:5}
\end{equation}
to the free Lagrangian $F^2_{ABC}$.  As in $D$=4,
the effect of this term is to shift the electric charges, 
but this time by
\begin{eqnarray}
e^{(i)'}_a & = & e^{(i)}_a - (3!) \,\theta 
\; \epsilon^{ij}g^{(j)}_a \; . 
\label{shift1}
\end{eqnarray}
The antisymmetry of this shift is traceable to
that of the $\theta$-term;  more explicitly, 
the $\theta$-term
involves only mixed couplings, with opposite signs:
$\theta A^{(1)}_{0n}
\partial_m B^{(2) n m}$, $-\theta A^{(2)}_{0n}
\partial_m B^{(1) n m}$. [Had we
taken opposite conventions for the couplings,
there would be a relative minus sign
in (\ref{quanti6D}) between the contributions of
the two fields, and the same sign in 
(\ref{shift1}).]  For the quantization condition
(\ref{quanti6D}) to be invariant under
the shift (\ref{shift1}) then requires
the $+$ sign there.  
Thus, also in $D$=6 a (generalized) 
$\theta$-angle argument 
determines the
sign in the quantization condition. 

From these two examples, it is clear that
the $(-)^p$ factor is a reflection of the opposite
symmetries of the $\theta$-terms $F^*\!F$ and
$\epsilon_{ij}F^{(i)}\,^*\!F^{(j)}$ in alternating
dimensions.
 
\noindent{\bf 2. ~Adjoining dimensions.}

We now turn to the argument from dimensional reduction
(actually, ``enhancement"). 
Since higher dimension is clearly 
more restrictive,  our logic will be to show
that the quantization rule in $D$=2$p$--2 for those specific
configurations obtained by reduction from $2p$ 
imposes the form of the $D$=2$p$ rule as well.
Specifically we shall show that if the
quantization condition holds with one sign in
$D$=2$p-$2, then it must hold with the opposite one
in $D$=2$p$; in particular, the $+$ sign in
$D$=6 follows from the $-$ sign in $D$=4.

We relate $D$=6 to $D$=4 by toroidal compactification, 
$M^6 = R^4 \times T^2$.  The spacetime coordinates 
$x^A$ ($A = 0, 1, 
\dots, 5$) split into $x^A = (x^\mu ; x^4, x^5)$ where
$(x^4,\;x^5)$ parametrize the torus which, for our
purposes, may be assumed to be the standard 
$(dx^4)^2 + (dx^5)^2$, with $(x^4, \; x^5)$
having respective ranges $[0,L_4]$ and $[0,L_5]$. [For
fields independent of $(x^4, \;x^5)$ as considered here,
one may always diagonalize the internal metric, 
but we chose {\it not} to also rescale the ranges to unity.]
The full spatial $5D$ rotational symmetry is broken by the 
compactification of course.  However, there is a useful
residual ``$Y$-symmetry," under simultaneous
interchange of $x_4/L_4$ with $x_5/L_5$ together with a
$4D$ parity (P) transformation. Indeed, we will conclude 
generally that the 
quantization condition in $D$=$2p$--2, together with 
$Y$-symmetry, implies the corresponding one at $D$=2$p$.

A non-chiral $2$-form $A_{AB}$ in $D$=6 induces
two $D$=4  U(1) gauge fields $A^{(i)}_\mu$.  
The reduction proceeds by assuming $A_{AB}$
to be constant along the internal tori and to have only 
$A_{4 \mu}$ and $A_{5 \mu}$ as non-zero components.  
[The other, $A_{\mu \nu}$ components and the higher 
modes induce further four-dimensional
fields which are not relevant to our discussion.]
The correspondence is
\begin{equation}
A^{(1)}_\mu = \sqrt{L_4 L_5} A_{4 \mu}, \; 
A^{(2)}_\mu = \sqrt{L_4 L_5} A_{5 \mu}
\label{correspondence}
\end{equation}
as follows from reduction of the $2$-form
action $\int d^6x F^2_{ABC}$.  In $D$=4 terms, (besides the P)
$Y$-transformations interchange the two $A^{(i)}$.

The $D$=6 sources that correspond to point particles 
in $D$=4
are strings winding around the internal torus
directions.  
For a single string along $x^4$
at ${\bf x}= 0$, $x^5 = a$, the
current has as its only non-vanishing components 
\begin{equation}
J^{04}_e = e \delta^{(3)}({\bf x}) \delta(x^5-a),\;\;\;
J^{04}_m =  g \delta^{(3)}({\bf x}) \delta(x^5-a)
\label{6Dcurrents0}
\end{equation}
where $(e, g)$ are the respective electric and magnetic
strengths of the string.  The zero modes of the $2$-form
field couple only to the zero modes of the source.
Thus, from the point of view of the
zero modes, one can replace the source by a continuous
distribution of parallel strings aligned along $x^4$, 
with constant electric and magnetic strengths
per unit length,  $(\rho_5,\sigma_5)$, along the transverse 
$(x^5)$
direction.  Such a distribution yields a membrane
wrapping around the torus and does not excite
the higher modes (``vertical reduction" of \cite{Stelle}).
This alternative description preserves translation
invariance along $x^5$. Replacing the above source by 
a stack of strings
at ${\bf x}= 0$ aligned along $x^4$
amounts to replacing the currents of (\ref{6Dcurrents0} )
by
\begin{equation}
J^{04}_e =  \rho_5 \delta^{(3)}({\bf x}),\;\;\;
J^{04}_m =  \sigma_5 \delta^{(3)}({\bf x}) \; .
\label{6Dcurrents}
\end{equation}
These currents are obtained by summing 
the currents of the individual
strings, {\it e.g.}, $J^{04}_e({\bf x},x^5)
= \rho_5 da \delta^{(3)}({\bf x}) \delta(x^5 - a)$ 
for the string located at
$x^5 = a$.  The corresponding $D$=6 charges are
\begin{equation}
e = \rho_5 L_5,\;\;\;\; g = \sigma_5 L_5.
\label{6Dstring1}
\end{equation}
From the $4D$ point of view, the stack appears to
have the $U(1)$ charges
\begin{equation}
\left( e^{(1)}, \, g^{(1)}, \, e^{(2)}, \, g^{(2)}\right) =
\left( e \sqrt{\frac{L_4}{L_5}}, 0, 0, 
 g \sqrt{\frac{L_4}{L_5}}\: \right)
\label{4Dstring1}
\end{equation}
as shown by  the analysis of the equations
of motion given below.
Again, we adopt the same sign conventions for the
two $U(1)$'s and define electric and magnetic charges in
$4D$ in such a way that 
$\mbox{\boldmath $\nabla$}\cdot {\bf E}^{(i)}\sim + e^{(i)}$, 
$\mbox{\boldmath $\nabla$}\cdot {\bf B}^{(i)}\sim + g^{(i)}$ 
(with same $+$ sign for both $i$).  Similarly, the 
current of a single dyonic string $(e',g')$
lying on the $x^5$ axis at $({\bf x}= {\bf b}, x^4 = c)$ 
is given by
\begin{equation}
J^{05}_e = e' \delta^{(3)}({\bf x-b}) \delta(x^4-c),\;\;\;
J^{05}_m =  g' \delta^{(3)}({\bf x-b}) \delta(x^4-c) \; .
\label{6Dcurrents0bis}
\end{equation}
Again, from the zero mode point of view, this can be replaced 
by a suitable stack
of strings whose
$6D$ currents are
\begin{equation}
J^{05}_e =  \rho_4 \delta^{(3)}({\bf x}- {\bf b}), \;\;\;
J^{05}_m = \sigma_4 \delta^{(3)}({\bf x}- {\bf b}) \; .
\label{6Dcurrents'}
\end{equation}
Here the $D$=6 charges are 
\begin{equation}
e' = \rho_4 L_4, \;\;\; g' = \sigma_4 L_4  \; ,
\label{6Dstring2}
\end{equation}
while the $4D$ charges are
\begin{equation}
\left( e^{(1)}, \, g^{(1)}, \, e^{(2)}, \, g^{(2)}\right)
=
\left( 0, - g' \sqrt{\frac{L_5}{L_4}}, 
e' \sqrt{\frac{L_5}{L_4}},0 \right) \;.
\label{4Dstring2}
\end{equation}

These  charges have two  properties:  First,
a dyonic string in $D$=6 along $x^4$ or $x^5$
does not appear as a dyon in $D$=4.  Rather, it is
{\it electrically} charged for one $U(1)$ and 
{\it magnetically} charged for the other $U(1)$.  To get dyons
for the same $U(1)$ in $D$=4, one needs to superpose
strings along both $x^4$ and $x^5$.  The same remark
applies to the chiral case (in $D$=6), for which the
two $D$=4 $U(1)$'s are related by the duality rotation 
${\bf B}^{(2)} =  {\bf E}^{(1)}$, ${\bf E}^{(2)} =
- {\bf B}^{(1)}$.  The above strings would appear as 
either purely electric 
(first string) or purely magnetic (second string) but
do not carry both types of charges.
Second, there is a crucial flip of sign in the magnetic
charges for the two $U(1)$'s. 
The simultaneous existence of the ``dual"
configurations (15) and (21) reflects the $Y$
symmetry. Indeed, a $Y$-transformation, to a $D$=4 observer,
just induces this dual exchange.
To understand how the $D$=4 assignments arise, 
consider 
the field equations, 
$\partial_A F^{ABC} = J^{BC}_e$, $\partial_A \, ^* F^{ABC}
= J^{BC}_m$, for the given sources 
in terms of the $D$=4 fields.  For the source
(\ref{6Dstring1}) along $x^4$,
the equations reduce to 
\begin{equation}
\partial_i E^{i(1)} \equiv 
+\sqrt{L_4 L_5} \partial_i F^{i04} 
= + e \sqrt{\frac{L_4} {L_5}} \delta^{(3)}
({\bf x})
\end{equation}
and 
\begin{equation}
\partial_i B^{i(2)} \equiv
+ \sqrt{L_4 L_5} \partial_i ((1/2!) \epsilon^{imn045} F_{mn5})
\equiv
+\sqrt{L_4 L_5} \partial_i \, ^* \! F^{i04}
= + g \sqrt{\frac{L_4} {L_5}} \delta^{(3)}
({\bf x}),
\end{equation}
where $i,m,n =1,2,3$.
For the source (\ref{6Dstring2}), one finds 
\begin{equation}
\partial_i E^{i(2)}  \equiv 
\sqrt{L_4 L_5}\partial_i F^{i05} 
= + e' \sqrt{\frac{L_5} {L_4}} \delta^{(3)} ({\bf x}-
{\bf b})
\end{equation}
and 
\begin{equation}
\partial_i B^{i(1)} \equiv
+ \sqrt{L_4 L_5} \partial_i ((1/2!) \epsilon^{imn054} 
F_{mn4})
\equiv - \sqrt{L_4 L_5} \partial_i \, ^* \! F^{i05}
= - g' \sqrt{\frac{L_5} {L_4}}
\delta^{(3)} ({\bf x}-{\bf b}),
\end{equation}
with a minus
sign because
$\epsilon^{imn054} = - \epsilon^{imn045}$.
This leads to the assignments (\ref{4Dstring1}) and 
(\ref{4Dstring2}). 

We now deduce the quantization
condition in $D$=6  from that in $D$=4. For the strings 
(\ref{4Dstring1}) and (\ref{4Dstring2}),
the $D$=6 quantization condition is
\begin{equation}
e g' \pm e' g = 2 \pi \hbar n, \; \; n \in Z. 
\label{6D}
\end{equation}
where we have again left the relative sign open. The 
quantization condition in $D$=4, on the
other hand, is, in terms of $D$=4 charges,
\begin{equation}
(e_a^{(1)} g_b^{(1)} -e_b^{(1)} g_a^{(1)})
 + (e_a^{(2)} g_b^{(2)} - e_b^{(2)} g_a^{(2)})
 =2 \pi \hbar n, \; \; n \in Z \; .
 \label {4D}
 \end{equation}
Recall that the relative $+$ sign between the 
two $U(1)$ contributions is due to our
identical coupling conventions for both.
The only choice that makes (\ref{4D}) consistent
with (\ref{6D}) is the $+$ sign as is easy to verify by 
using the explicit 
values of the $D$=4 charges in terms of the $D$=6 ones.
To show, finally, that the electric and magnetic charges
of a single string in $D$=6 are constrained by
$ 2 e g= 2 \pi n \hbar , \; \; n \in Z $, we recall that this
condition was obtained in \cite{DGHT1} 
by exploiting the flexibility of Dirac membranes to 
perform motions that do not distinguish between the
spatial directions.  It comes  as no surprise therefore, 
that one can recover this relation from the $D$=4 
point of view by using $Y$ symmetry.
Indeed, together with the configuration $(e,0,0,g)$, 
$Y$ implies that the configuration $(0,-g,e,0)$ should also
exist. Applying
the $D$=4 dyon quantization condition to
the two configuration appearing above,
we recover this $e-g$ relation. We can imagine continuing 
this chain of 
arguments inductively: (1)~
consider the dyonic configuration in $2p$ dimensions;
(2)~ list the $2p-2$ dimensional configurations 
to which it gives rise, including those related by
$Y$-symmetry; (3)~ apply
the $2p-2$ dimensional quantization rules which 
will therefore relate the $2p$ dimensional parameters,
etc. So starting with say
the $D$=4 quantization rules, those for higher 
dimensions will follow, and it is clear that there is 
a $(-)^p$ alternation.

In retrospect, it is not surprising that one can infer the
$D$=6 quantization condition from that in $D$=4, together
with the extra $Y$-symmetry it enjoys.  Indeed, as was
shown in \cite{DGHT2}, the respective quantization
conditions with $+/-$ signs possess exactly the same general
solutions (assuming existence of pure electric sources);
hence (when  $(C)P$ invariance is imposed in $D$=4) 
they are clearly equivalent  \cite{Witten}.

To summarize, we have provided two independent
derivations of the $(-)^p$ sign factor in the $p$-brane
dyon quantization conditions.  Both arguments are
ultimately manifestations of the basic ``double dual"
identity $**=(-)^p$.

\section*{Acknowledgements}

A.S.\ gratefully acknowledges very stimulating discussions with 
D. Kutasov; S.D.\ and M.H.\ thank A.\ Gomberoff and
C.\ Teitelboim. The work of S.D.\ was supported 
by NSF grant PHY 93-15811, that of
M.H.\ by I.I.S.N.\  (Belgium), and that of A.S.\ by
the Israeli Academy of Sciences and the Minerva Foundation
(Germany).  M.H.\ and A.S.\ thank Carmen Nu\~ nez 
and the Bariloche High Energy Group for hospitality 
at the Bariloche Institute, where this work began.

\end{document}